\documentclass[a4paper,11pt]{article}

\usepackage{amsmath}
\usepackage{latexsym}
\usepackage{amssymb}
\usepackage{amsfonts}
\usepackage[dvips]{graphicx}

\numberwithin{equation}{section}

\DeclareMathOperator{\Jac}{tridiag}
\newcommand{\E}{ \, {\mathsf E}\, }
\newcommand{\C}{ \mathbb{C} }
\newcommand{\R}{ \mathbb{R} }
\newcommand{\Z}{ \mathbb{Z} }
\newcommand{\re}{\mathop{\mathrm{Re}}}
\newcommand{\im}{\mathop{\mathrm{Im}}}

\newcommand{\s}{\mathop{\mathrm{spec}}}
\newcommand{\Supp}{\mathop{\mathrm{Supp}}}
\newcommand{\dist}{\mathop{\mathrm{dist}}}
\newcommand{\diag}{\mathop{\mathrm{diag}}}

\newtheorem{Th}{Theorem}[section]
\newtheorem{Lemma}[Th]{Lemma}
\newtheorem{Prop}[Th]{Proposition}
\newtheorem{Corol}[Th]{Corollary}

\author{Ilya Ya. Goldsheid and Boris A. Khoruzhenko\\
        School of Mathematical Sciences, Queen Mary,
        University of London,\\ London E1 4NS, U.K.}

\title{Eigenvalue curves of asymmetric
tridiagonal random matrices}

\date{26 October 2000}

\begin{document}

\maketitle
\begin{abstract}
Random Schr\"odinger operators with imaginary vector potentials
are studied in dimension one. These operators are non-Hermitian
and their spectra lie in the complex plane. We consider the
eigenvalue problem on finite intervals of length $n$ with periodic
boundary conditions and describe the limit eigenvalue distribution
when  $n\to \infty$. We prove that this limit distribution is
supported by curves in the complex plane. We also obtain equations
for these curves and for the corresponding eigenvalue density in
terms of the Lyapunov exponent and the integrated density of
states of a ``reference'' symmetric eigenvalue problem. In
contrast to these results, the spectrum of the limit operator in
$l^2({\bf Z})$ is a two dimensional set which is not approximated
by the spectra of the finite-interval operators.

\bigskip


\bigskip


\end{abstract}

\vspace{6ex}


\section{Introduction}
\label{section1}

Consider an infinite asymmetric tridiagonal matrix $J$ with real
entries $q_k$ on the main diagonal and positive entries $p_k$ and
$r_k$ on the sub- and super-diagonal. Cut a square block of size
$n=2m+1$ with the center at $(0,0)$ out of $J_n$ and impose the
periodic boundary conditions in this block. The obtained matrix
has the following form
\begin{equation}\label{000}
J_n = \left(
\begin{array}{l l l l l }
q_{-m}  & r_{-m}  &     &          & p_{-m}
\\[2ex]
p_{-m+1} &  q_{-m+1}  & r_{-m+1} & &
\\
  & \hspace{-5ex} \ddots    & \hspace{-5ex}  \ddots   &
\hspace{-5ex} \ddots &
\\
  & p_{0} &  q_{0}  & r_0 &
\\
   & \hspace{3ex} \ddots    & \hspace{3ex}  \ddots   &
\hspace{3ex} \ddots &
\\
     & &p_{m-1} &q_{m-1} &r_{m-1}
\\[2ex]
r_m  &    & &p_m &q_m
\end{array}
\right)
\end{equation}
We show nonzero entries of $J_n$ only.

When $n$ is large, the spectrum of $J_n$ cannot be obtained
analytically, except for some special choices of $p_k$, $q_k$, and
$r_k$. However, it can be easily computed for ``reasonable''
values of $n$ using any of the existing linear algebra software
packages. If the $p_k$, $q_k$, and $r_k$ are chosen randomly the
results of such computations are striking. For large values of
$n$, the spectra of $J_n$ lie on smooth curves which change little
from sample to sample. This was observed by Hatano and Nelson
\cite{HN1,HN2}.

\begin{figure}[ht]
\label{fig1} \centerline{
\includegraphics[width=14cm]{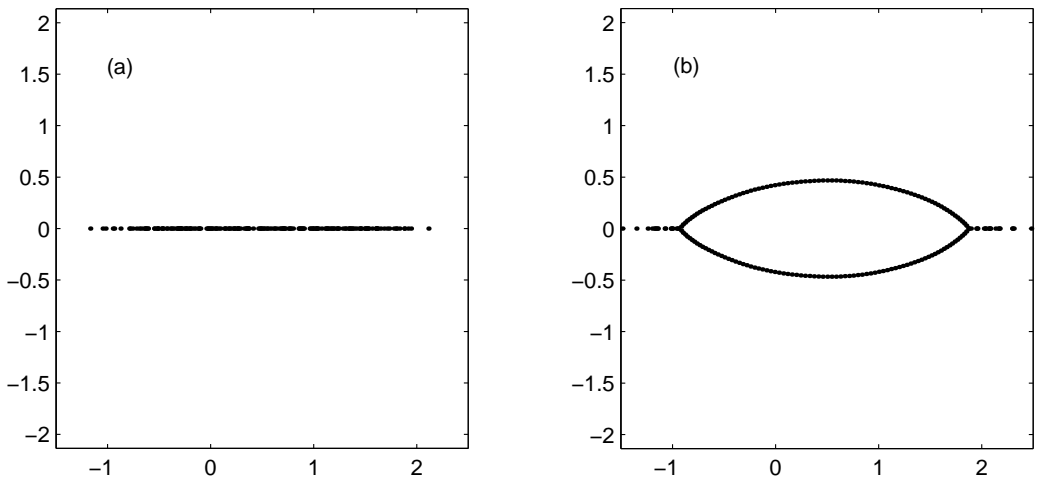}
 }
 \caption{Spectra of $J_n$ ($n=201$) where (a) all non-zero
entries are drawn from Uni$[0,1]$; and (b) the sub-diagonal and
diagonal entries are drawn from Uni$[0,1]$ and super-diagonal
entries are drawn from Uni$[\frac{1}{2},1\frac{1}{2}]$.}
\end{figure}
Fig.\ \ref{fig1} shows spectra of two matrices $J_n$ of dimension
$n=201$. One matrix has its non-zero entries drawn from
Uni$[0,1]$\footnote{Uni$[a,b]$ denotes the uniform distribution on
$[a,b]$.}. Its spectrum is shown on plot (a). Note that this
matrix is only \emph{stochastically} symmetric. For a typical
sample from Uni$[0,1]$, $J_n\not=J^T_n$. Nevertheless its spectrum
is real. The other matrix has its diagonal and sub-diagonal
entries drawn from Uni$[0,1]$ and super-diagonal entries drawn
from Uni$[\frac{1}{2},1\frac{1}{2}]$. Its spectrum is shown on
plot (b).

\begin{figure}[ht]
\label{fig2} \centerline{
\includegraphics[width=14cm]{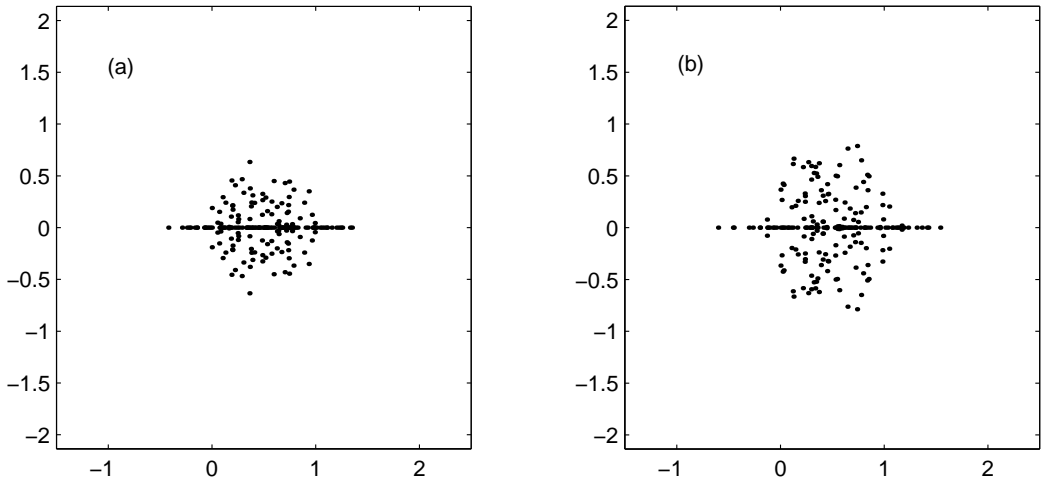}
 }
 \caption{Spectra of $J_n$ ($n=201$) where (a) the sub- and super-diagonal
 entries are drawn from Uni$[-\frac{1}{2},\frac{1}{2}]$ and the diagonal
 entries are drawn from Uni$[0,1]$; and (b) the sub-diagonal
 entries are
 drawn from Uni$[-\frac{1}{2},\frac{1}{2}]$, and the diagonal and super-diagonal
 are drawn from Uni$[0,1]$}
\end{figure}

Fig.\ \ref{fig1} is in a sharp contrast to our next figure. We
took the two matrices of Fig.\ \ref{fig1} and subtracted
$\frac{1}{2}$ from all their sub- and super-diagonal entries,
including the corner ones. Fig.\ \ref{fig2} shows spectra of the
obtained matrices. Note that these spectra have a two-dimensional
distribution. As will soon become clear, the eigenvalue curves in
Fig.\ \ref{fig1} are due to the sub- and super-diagonal entries
having the same sign.

The class of random matrices (\ref{000}) was introduced by Hatano
and Nelson in 1996 \cite{HN1,HN2}. Being motivated by statistical
physics of magnetic flux lines and guided by the relevant physical
setup, they considered random non-Hermitian Schr\"odinger
operators $H(g)=(i\frac{d}{dx} +ig)^2 +V$ and their discrete
analogues $J_n$ in a large box with periodic boundary conditions
and discovered an interesting localization -- delocalization
transition. Hatano and Nelson also argued that the eigenvalues
corresponding to the localized states are real and those
corresponding to the delocalized states are non-real. Since then
there has been considerable interest to the spectra of $J_n$ and
their multi-dimensional versions in the physics literature.

Analytic descriptions of the spectrum of the random matrices
$J_n$, in the limit $n\to\infty$, were obtained independently and
almost simultaneously in \cite{BSB,BZ} and in our paper \cite{GK}.
In this paper we present complete proofs of the results stated in
\cite{GK}. Our results do not rely on a particular asymptotic
regime (as in \cite{BSB}) or a particular probability distribution
(see \cite{BZ}). We explain the phenomenon of eigenvalue curves
depicted in Fig.\ \ref{fig1}. Under general assumptions on the
probability distribution of the $p_k$, $q_k$, and $r_k$,  we prove
that the eigenvalues of $J_n$ tend to certain non-random curves in
the complex plane when $n\to\infty$. We also obtain an analytic
description for these curves and for the corresponding limiting
distribution of eigenvalues. This limiting eigenvalue distribution
may undergo a transition from a distribution on the real line, as
in Fig. \ref{fig1}(a), to a distribution in the complex plane, as
in Fig.\ \ref{fig1}(b), when parameter values of the probability
laws of the matrix entries vary.

Our results resemble those obtained  in the 1960's for truncated
asymmetric Toeplitz matrices \cite{Schmidt,Hirschman}. The
eigenvalues of finite blocks of an infinite Toeplitz matrix are
distributed along curves in the complex plane in the limit when
the block size goes to infinity, see \cite{Widom2} for a survey.
Of course, this resemblance is only formal. The class of matrices
we consider is very different from Toeplitz matrices.

It is interesting that the spectra of the finite matrices $J_n$
with random entries, even in the limit $n\to\infty$, are entirely
different from the spectrum of the corresponding infinite random
matrix $J=\Jac (p_j,q_j,r_j)$ considered as an operator acting on
$l^2(\Z)$. Indeed, suppose that the non-zero entries of $J$ are
independently drawn from a three-dimensional distribution with a
bounded support $S$. Then, the probability is one that for any
given $(p,q,r)\in S\subset \mathbb{R}^3$ and for any $n\in
\mathbb{N}$ and $\varepsilon>0$ we can always find in $J$ a block
of size $n$ such that for all $j$ within this block
$|p_j-p|<\varepsilon$, $|q_j-q|<\varepsilon$, and
$|r_j-r|<\varepsilon$. By the Weyl criterion, this implies that
with probability one the spectrum of $J=\Jac (p_j,q_j,r_j)$
contains the spectra of $\Jac (p,q,r)$ for all $(p,q,r)\in S$.
(This argument is well known in the spectral theory of random
operators, see e.g.\ \cite{CL,PF}.) Since for every $(p,q,r)$ the
spectrum of $\Jac (p,q,r)$ is the ellipse
\[
\{z: z=pe^{it}+q+r^{-it}, \hspace{1ex} t\in [0,2\pi]\},
\]
the spectrum of the infinite tridiagonal random matrix $J$ has a
two-dimensional support with probability one if $S$ is suffiently
rich. That is why the
eigenvalue curves of $J_n$ are surprising. This discrepancy
between the spectra of $J_n$ and $J$ was mentioned in the
preliminary account of our work \cite{GK} and was one of our
motivations for studying the eigenvalue distribution for $J_n$.
Nothing of this kind happens for periodic sequences
$\{(p_j,q_j,r_j)\}$. If $\{p_k\}$, $\{q_k\}$ and $\{r_k\}$ have a
common period, then the spectrum of $J$ lies on a curve in the
complex plane \cite{Naiman} coinciding with the eigenvalue curves
of $J_n$ in the limit $n\to \infty$. The two-dimensional spectra
are specific to ``sufficiently rich'' random non-selfadjoint
operators (see recent works \cite{D,D1} which contain a much more
detailed analysis of the spectral properties of infinite
tridiagonal random matrices).

One significant consequence of the above mentioned  discrepancy
between the spectra of $J_n$ and $J$ is that the norm of the
resolvent $(J_n - zI_n)^{-1}$ may tend to infinity as $n\to\infty$
even when $z$ is separated from the spectra of $J_n$. This aspect
of instability inherent in non-normal matrices \cite{T1,T2} was
thoroughly examined for asymmetric Toeplitz matrices in \cite{RT}
and for random bidiagonal matrices in \cite{T}. We do not discuss
it here.

The rest of this paper is organized as follows. Our main results,
Theorems \ref{th1} and \ref{th2}, are stated in Section
\ref{section1a}. We also discuss there several
corollaries of our theorems. Theorem \ref{th1} is proved in Section
\ref{section2} and Theorem \ref{th2} is proved in Section
\ref{section3}. Our proofs use a number of results from
the spectral theory of (selfadjoint) random operators. These are
summarized in Appendix.

\bigskip

{\bf Acknowledgements} We thank E.B. Davies and L.N.
Trefethen for very useful and interesting discussions, and
M. Sodin for bringing our attention to paper \cite{Naiman}.

\section{Main results and corollaries}
\label{section1a}

To simplify the notation, we label entries of $J_n$ by $(j,k)$
with $j$ and $k$ being integers between 1 and $n$. We also set
$p_k=-e^{\xi_{k-1}}$ and $r_k=-e^{\eta_{k}}$, so that $J_n$ takes
the form
\begin{equation}\label{3}
J_n = \left(
\begin{array}{c c c c}
q_{1}  & -e^{\eta_{1}}   &            & -e^{\xi_{0}}
\\
-e^{\xi_{1}} &  \hspace{-2ex}\ddots  & \hspace{-2ex}\ddots     &
\\
       &\hspace{-2ex} \ddots    &  \hspace{-2ex} \ddots   & \hspace{-2ex} -e^{\eta_{n-1}}
\\
-e^{\eta_{n}}     &          &  \hspace{-2ex}-e^{\xi_{n-1}}  &
\hspace{-2ex} q_n
\\
\end{array}
\right),
\end{equation}
where only the non-zero entries of $J_n$ are
shown\footnote{Following the tradition, we have chosen the sub-
and super-diagonal entries to be negative. Provided the sub- and
super-diagonal entries are of the same sign, the choice of sign
does not matter.}. The corresponding eigenvalue equation can be
written as the second-order difference equation
\begin{equation}\label{1}
-e^{\xi_{k-1}}\psi_{k-1} -e^{\eta_{k}} \psi_{k+1} +q_{k}\psi_{k} =
z\psi_{k},  \;\;\; 1\le k \le n,
\end{equation}
with the boundary conditions (b.c.)
\begin{equation}\label{2}
\psi_{0}=\psi_{n}\ \mathrm{and}\  \psi_{n+1}=\psi_{1}.
\end{equation}
Our basic assumptions are:--
\begin{eqnarray}\label{ba1}
& &
\begin{array}{l}
\{ (\xi_k, \eta_k, q_k) \}_{k=0}^{\infty} \hbox{ is a stationary
ergodic (with respect to translations $k \longrightarrow k+1$) }
\\ \hbox{sequence of 3-component random vectors defined on a common probability }
\\ \hbox{space  $(\Omega, {\cal F}, P)$};
\end{array}\hspace{5ex}
\\
\label{ba2} & &
\begin{array}{l}
\E \ln (1+|q_0|),\ \E \xi_0,\ \E \eta_0\ \hbox{are finite.}
\end{array}
\end{eqnarray}
The symbol $\E$ stands for the integration with respect to the
probability measure, $\E f= \int_{\Omega} f dP$.


We have chosen the off-diagonal entries of $J_n$ to be of the same
sign. More generally, one could consider matrices whose
off-diagonal entries satisfy the following condition:
\begin{itemize}
\item[(*)] the product of $(k,k+1)$ and $(k+1,k)$ entries is
positive for all $k=1,2, \ldots n-1$.
\end{itemize}
Any real asymmetric \emph{purely} tridiagonal matrix satisfying
(*) can be transformed into a symmetric tridiagonal matrix. The
recipe is well known: put $\psi_k=w_k\varphi_k$ in (\ref{1}) and
choose the $w_k$ so that to make the resulting difference equation
symmetric. This is always possible when (*) holds, and the weights
$w_k$ are defined uniquely up to a multiple.

The above transformation is the starting point of our analysis. We
set $w_0=1$ and
\begin{eqnarray}\label{d}
 w_k=e^{ \frac{1}{2} \sum_{j=0}^{k-1} (\xi_j-\eta_j) },\;
k\ge 1.
\end{eqnarray}
Eqs. (\ref{1})--(\ref{2}) are then transformed into
\begin{equation}\label{4}
-c_{k-1}\varphi_{k-1}-c_{k} \varphi_{k+1}
+q_{k}\varphi_{k}=z\varphi_{k},\;\;\;\;\;\;    1\le k \le n,
\end{equation}
\begin{equation}
\label{5} \varphi_{n+1}=\frac{w_1}{w_{n+1}}\varphi_{1},
\hspace{2ex} \varphi_{n}=\frac{1}{w_n}\varphi_{0},
\end{equation}
where
\begin{equation}\label{c_}
c_k= e^{\frac{1}{2}(\xi_k+\eta_k)}.
\end{equation}

In the matrix form the transformation from $\psi$ to $\varphi$ can
be expressed as $J_n \mapsto W^{-1}J_nW$ with $W=\diag (w_1,
\ldots w_n)$. Eqs.\ (\ref{4}) -- (\ref{5}) can be rewritten as
$W^{-1}J_nW \varphi=z\varphi$, where $\varphi=(\varphi_1, \ldots ,
\varphi_n)^T$. The asymmetry in the transformed eigenvalue problem
is due to boundary conditions (\ref{5}). One can visualize
this asymmetry: obviously,
\begin{equation}\label{simtr}
W^{-1}J_nW=H_n+V_n,
\end{equation}
\begin{equation}\label{7}
H_n = \left(
\begin{array}{c c c c}
q_{1}  & -c_{1}   &            &
\\
-c_{1} &  \hspace{-2ex}\ddots  & \hspace{-2ex}\ddots     &
\\
       &\hspace{-2ex} \ddots    &  \hspace{-2ex} \ddots   & \hspace{-2ex} -c_{n-1}
\\
       &          &  \hspace{-2ex}-c_{n-1}  & \hspace{-2ex} q_n
\\
\end{array}
\right)\ \hbox{and}\;\; V_n= \left(
\begin{array}{c c c c c c}
      &           &            &  & &
a_n
\\
 &     &      & &
\\
 &     &      & & &
\\
       &     &    & & &
\\
b_n   &          &     & & &
\\
\end{array}
\right),
\end{equation}
where
\begin{equation}\label{ab}
a_n = -c_0w_n = - c_0e^{\frac{1}{2}\sum_{j=0}^{n-1}
(\xi_j-\eta_j)} \hspace{2ex} \hbox{and}\hspace{2ex} b_n
=-c_n\frac{w_1}{w_{n+1}}= -c_ne^{-\frac{1}{2}\sum_{j=1}^n
(\xi_j-\eta_j)}.
\end{equation}
$H_n$ is a real symmetric Jacobi matrix. Its eigenvalue equation
is given by Eq.\ (\ref{4}) with the Dirichlet b.c.\
\begin{equation}
\label{2b} \psi_{n+1}=\psi_0=0.
\end{equation}
$V_n$, which is due to (\ref{5}), is a real asymmetric matrix. If
$\E (\xi_0 - \eta_0 )\not=0$ then one of the two non-zero entries
of $V_n$ increases, and the other decreases exponentially fast
with $n$ (with probability 1).

Though one could deal directly with $J_n$, we deal with $H_n+V_n$
instead. We thus consider the asymmetric eigenvalue problem
(\ref{4}) -- (\ref{5}) as an exponentially large ``perturbation''
of the symmetric problem \ (\ref{4}), (\ref{2b}). This
point of view allows us to use the whole bulk of information about
the symmetric problem (\ref{4}), (\ref{2b}) in the context of the
asymmetric problem (\ref{4})-(\ref{5}). It is worth mentioning
that the exponential rate of growth of $V_n$ is very essential: no
interesting effects would be observed for sub-exponential rates.

Following the standard transfer-matrix approach, we rewrite
(\ref{4}) as
\begin{equation}\label{10}
\left(
\begin{array}{l}
\varphi_{k+1} \\ \varphi_k   \\
\end{array}
\right) = A_k \left(
\begin{array}{l}
\varphi_{k} \\ \varphi_{k-1}   \\
\end{array}
\right), \;\;\;\;\hbox{where}\ A_k =\frac{1}{c_k} \left(
\begin{array}{l l}
q_{k}-z  & -c_{k-1}    \\ c_k      &  \phantom{-} 0    \\
\end{array}
\right),
\end{equation}
and introduce the notation
\begin{equation}\label{tranmat}
S_n (z) = A_n\cdot \ldots \cdot A_1.
\end{equation}

In order to formulate our results, we need to recall the classical
notion of the Lyapunov exponent associated with Eq.\ (\ref{4}):
\begin{equation}\label{13}
\bar \gamma (z)=\lim_{n\to \infty} \frac{1}{n} \E \log ||S_n
(z)||.
\end{equation}
It is well known that the limit in (\ref{13}) exists for every
$z\in \C$ and is non-negative. Obviously any matrix norm can be
used in (\ref{13}), as they all are equivalent. It is convenient
for our purposes to use the following norm
\begin{equation}\label{norm}
||M||=\max_{k}\sum_{j}{|M_{jk}|}.
\end{equation}

Introduce
\begin{equation}\label{g}
g=\frac{1}{2}\E(\eta_0-\xi_0)
\end{equation}
and consider the curve
\begin{equation}\label{14}
{\cal L}= \{z\in \C: \hspace{3ex} \bar \gamma (z) = |g|
\hspace{1ex} \}.
\end{equation}
This curve separates the two domains
\begin{equation}\label{14a}
D_1= \{z\in \C: \hspace{3ex} \bar \gamma (z) > |g| \hspace{1ex} \}
\hspace{3ex}\hbox{and} \hspace{3ex} D_2= \{z\in \C: \hspace{3ex}
\bar \gamma (z) < |g|\hspace{1ex} \}
\end{equation}
in the complex plane. Note that $D_2$ may be empty for some values
of $g$. In this case, $\cal L$ is either empty as well or
degenerates into a subset of $\R$. In Section \ref{section2} we
prove the following

\bigskip

\begin{Th}\label{th1} Assume (\ref{ba1}) -- (\ref{ba2}). Then
for $P$-almost all $\omega=\{(\xi_j,\eta_j,q_j) \}_{j=0}^{\infty}$
the following two statements hold:
\begin{itemize}
\item[(a)] For every compact set $K_1\subset D_1\backslash\R$ there exists an
integer number $n_1(K_1, \omega)$ such that for all $n> n_1(K_1,
\omega)$ there are no eigenvalues of $J_n$ in $K_1$.
\item[(b)] For any compact set $K_2\subset D_2$ there exists an integer number $n_2(K_2,
\omega)$ such that for all $n> n_2(K_2, \omega)$ there are no
eigenvalues of $J_n$ in $K_2$.
\end{itemize}
\end{Th}

\bigskip

To proceed, we need to introduce another well studied function,
the integrated density of states $N(\lambda)$ associated with Eq.\
(\ref{4}). Let
\[
N_n(\lambda) = \frac{1}{n}\# \{\hspace{0.5ex}\hbox{eigenvalues of
$H_n$ in $(-\infty; \lambda)$} \hspace{0.5ex} \}.
\]
Then
\begin{equation}\label{N}
N(\lambda )=\lim_{n\to\infty} N_n(\lambda).
\end{equation}
It is well known that under assumptions (\ref{ba1}) --
(\ref{ba2}), the limit in Eq.\ (\ref{N}) exists on a set of full
probability measure and that $N(\lambda )$ is a non-random
continuous function, see e.g.\ \cite{CL,PF}.

It is a fact from spectral theory of random operators that $\bar
\gamma (z)$ and $N(\lambda)$ are related via the Thouless formula
\cite{CL,PF}
\begin{equation}\label{18}
\bar \gamma (z)=\int_{-\infty}^{+\infty} \log|z-\lambda |\
dN(\lambda) - \E\log c_0.
\end{equation}
According to this formula, $\bar \gamma (z)$, up to  the additive
constant $\E \log c_0=\frac{1}{2}\E(\xi_0 +\eta_0)$, is the
log-potential of $dN(\lambda )$:
\begin{equation}\label{018}
\Phi(z)=\int_{-\infty}^{+\infty} \log|z-\lambda |\  dN(\lambda).
\end{equation}
Then ${\cal L}$ is an equipotential line:
\begin{equation}\label{0018}
{\cal L}=\{z\in \C: \hspace{3ex} \Phi (z)=\max (\E\xi_0,
\E\eta_0)\hspace{1ex} \}.
\end{equation}
This equipotential line consists typically of closed contours
${\cal L}_j$. In turn, each contour consists of two symmetric arcs
whose endpoints lie on the real axis. The arcs are symmetric with
respect to the reflection $z \mapsto\overline{z}$. The
domain $D_2$ defined above is simply the interior of the contours
${\cal L}_j$.

Part (b) of Theorem \ref{th1} states that for almost all $
\{(\xi_j,\eta_j,q_j) \}_{j=0}^{\infty}$ the spectrum of $J_n$ is
wiped out from the interior of every contour ${\cal L}_j$ as
$n\to\infty$. Parts (a) and (b) together imply that for $P$-almost
all $ \{(\xi_j,\eta_j,q_j) \}_{j=0}^{\infty}$ the eigenvalues of
$J_n$ in the limit $n\to\infty$ are located on $\cal L\cup\R$.

Our next result describes the limiting eigenvalue distribution on
${\cal L}\cup \R$. Let $d\nu_{\!\scriptscriptstyle{J_n}}$ denote
the measure in the complex plane assigning the mass $1/n$ to each
of the $n$ eigenvalues of $J_n$.

\bigskip

\begin{Th}\label{th2}
Assume (\ref{ba1}) -- (\ref{ba2}). Then for $P$-almost all
$\omega=\{(\xi_j,\eta_j,q_j) \}_{j=0}^{\infty}$ the following
statement holds: For every bounded continuous function $f(z)$
\begin{equation}\label{ld}
\lim_{n\to\infty}\int_{\C} f(z) d\nu_{\!\scriptscriptstyle{J_n}}
(z) = \int_{\Sigma} f(\lambda) dN(\lambda) + \int_{\cal L} f(z(l))
\rho (z(l)) dl,
\end{equation}
where
\begin{equation}\label{Sigma}
\Sigma = \{\lambda \in \R: \hspace{3ex} \lambda \in \Supp dN,
\hspace{1ex} \Phi (\lambda+i0)
> \max (\E\xi_0, \E\eta_0) \hspace{1ex} \},
\end{equation}
\begin{equation}
\label{aaa} \rho (z) =
\frac{1}{2\pi}\left|\int_{-\infty}^{+\infty}\frac{dN(\lambda
)}{\lambda -z} \right|, \hspace{6ex} z\notin \R,
\end{equation}
and $dl$ is the arc-length element on $\cal L$.
\end{Th}

\bigskip

This theorem is proved in Section \ref{section3}. Of course the
eigenvalue curve $\cal L$ and the density of eigenvalues $\rho
(z)$ on it can be found explicitly only in exceptional
cases\footnote{One such case \cite{GK,BZ} is when  $\xi_j\equiv g$
and  $\eta_j \equiv -g$ for all $j$, and the diagonal entries
$q_j$ are Cauchy distributed.}. However, one can infer from our
theorems rather detailed general information about the spectra of
$J_n$ in the limit $n\to\infty$.

To facilitate the discussion, let us replace our basic assumption
(\ref{ba1}) by the following more restrictive but still quite
general one:
\begin{equation}\label{ba3}
\begin{array}{l} \{ (q_k, \xi_k, \eta_k) \}_{k=0}^{\infty} \hbox{ is
a sequence of independent identically distributed random }
\\ \hbox{vectors defined on a common
probability space.\hspace{28ex}}
\end{array}
\end{equation}

Under assumptions (\ref{ba1}) and (\ref{ba3}), the Lyapunov
exponent $\bar \gamma (z)$ is continuous in $z$ everywhere in the
complex plane, see e.g.\  \cite{BL}. Also, there exist positive
constants $C_0$ and $x_0$ depending on the distribution of
$(\xi_j, \eta_j, q_j)$ such that for all $|x|>x_0$
\begin{equation}\label{estimate}
\log |x| -C_0 <\bar \gamma (x) < \log|x| +C_0.
\end{equation}
These inequalities are obvious if the law $F$ of distribution of
$(\xi_j,\eta_j,q_j)$ has bounded support. If the support of $F$ is
unbounded, then (\ref{estimate}) can be obtained using methods of
\cite{G}. The continuity of $\bar \gamma (z)$ together with
(\ref{estimate}) imply that:

\hspace{1.5ex}(a) ${\cal L}$ is not empty if and only if
$\min_{x\in R}\bar \gamma (x)\le |g|$;

\hspace{1.5ex}(b) ${\cal L}$ is confined to a finite disk of
radius $R$ depending on the distribution of $(\xi_j, \eta_j,
q_j)$.

\smallskip
 To describe $\cal L$ we notice that $\bar \gamma (x+iy)$
is a strictly monotone function of $y\geq 0$. This follows from
the Thouless formula (\ref{18}). Hence, if $\bar \gamma (x+iy)
=|g|$ then
\begin{equation}\label{zz}
\bar\gamma(x) \le |g|
\end{equation}
and, vice versa, for each $x$ satisfying (\ref{zz}) one can find
only one non-negative $y(x)$ such that $z=x+iy(x)$ solves the
equation
\begin{equation}\label{14aa}
\bar \gamma (z)=|g|.
\end{equation}
Because of the continuity of $\bar\gamma(x) $, the set of $x$
where (\ref{zz}) holds is a union of disjoint intervals
$[a_j,a_j^\prime]$ with $ a_j<a_j^\prime$. Therefore ${\cal L}$ is
a union of disconnected contours ${\cal L}_j$. Each ${\cal L}_j$
consists of two smooth arcs, $y_j(x)$ and $-y_j(x)$, formed by the
solutions of Eq.\ (\ref{14aa}) when $x$ is running over
$[a_j,a_j^\prime]$. Apart from the specified contours, the set of
solutions of Eq.\ (\ref{14aa}) may also contain real points. These
are the points where $\bar \gamma (x)=|g|$.

It is easy to construct examples with a prescribed finite number
of contours. However we do not know any obvious reason for the
number of contours to be finite for an arbitrary distribution of
$(\xi_j,\ \eta_j,\ q_j)$.

According to (\ref{ld}), the limiting eigenvalue distribution may
have two components. One, represented by the first term on the
right-hand side in (\ref{ld}), is supported on the real axis. We
call this component real. The other, represented by the second
term, is supported by $\cal L$. We call this component complex.

The following statements are simple corollaries of our Theorems.
Assume (\ref{ba1}) and (\ref{ba3}) and  consider
\emph{stochastically} symmetric matrices $J_n$, i.e.\
$\E\xi_j=\E\eta_j$. In this case $g=0$ and hence the curve $\cal
L$ is empty. Therefore the limiting eigenvalue distribution has
the real component only. This is surprising and does not seem to
be obvious a priori. But even more surprising is that $\cal L$
remains empty for all
\[
|g|< g^{(1)}_{\hbox{\tiny{cr}}}\equiv \min_{x\in \R} \bar \gamma
(x).
\]
If the distribution of $(\xi_j,\eta_j,q_j)$ is such that the
support of the marginal distribution of $q_j$ contains at least
two different points then $\bar \gamma (x )$ is strictly positive
for all $x\in \R$ \cite{CL,PF}. In this case, by the continuity of
$\bar \gamma (x )$ and (\ref{estimate}),
$g^{(1)}_{\hbox{\tiny{cr}}}
>0$.

On the other hand, if
\[
|g|> g^{(2)}_{\hbox{\tiny{cr}}}\equiv \max_{x\in \Supp dN} \bar
\gamma (x).
\]
then $\Sigma$ of (\ref{Sigma}) is empty and the limiting
eigenvalue distribution has the complex component only. Obviously,
if the law of distribution of $(\xi_j,\eta_j,q_j)$ has unbounded
support, then $g^{(2)}_{\hbox{\tiny{cr}}} = +\infty$.

If $ g^{(1)}_{\hbox{\tiny{cr}}} < |g| <
g^{(2)}_{\hbox{\tiny{cr}}}$, then the real and complex components
coexist.

\smallskip
It is worth mentioning that the density $\rho (z)$ of the non-real
eigenvalues, see (\ref{aaa}), is analytic everywhere on ${\cal L}$
except the (real) end-points of the arcs. (If the limit eigenvalue
distribution has no real component than $\rho (z)$ is analytic
everywhere on $\cal L$.) The behavior of $\rho (z) $ near an
end-point of an arc,  $a_j$ say, depends on the regularity
properties of $N(\lambda )$ at this point. If the density of
states $dN(\lambda )/d\lambda$ of the reference equation (\ref{4})
is smooth in a neighborhood of $\lambda =a_j$ then $\rho (z)$ has
a finite limit as $z$ approaches $a_j$ along the arc. If, in
addition, $a_j$  belongs to both $\Sigma$ and $\cal$ then the
tangent to the arc at $a_j$ exists and is not vertical. In other
words, if $dN(\lambda )/d\lambda$ is smooth in a neighborhood of a
branching point $\lambda = a_j $ the complex branches grow out of
this point {\em linearly}.

\section{Eigenvalue curves}
\label{section2}
In this section we prove Theorem \ref{th1}. According to
(\ref{simtr}) the eigenvalues of $J_n$ and  $H_n+V_n$ coincide. It
is more convenient for us to deal with $H_n+V_n$ and we thus
consider the eigenvalue problem (\ref{4})--(\ref{5}). By
(\ref{10}) -- (\ref{tranmat}), we may write
$
(\varphi_{k+1},\varphi_k)^T= S_k (z) (\varphi_1,\varphi_0)^T$,
$k=1,\ldots, n,$ instead of (\ref{4}). In particular,
\[
\left(
\begin{array}{l}
\varphi_{n+1} \\ \varphi_n   \\
\end{array} \hspace{-1ex}
\right) = S_n (z) \left(
\begin{array}{l}
\varphi_{1} \\ \varphi_{0}   \\
\end{array}
\right).
\]
On the other hand, as required by (\ref{5}),
\[
\left(
\begin{array}{l}
\varphi_{n+1} \\ \varphi_n   \\
\end{array}\hspace{-1ex}
\right) = \frac{1}{w_n} \left(\hspace{-0.5ex}
\begin{array}{c c}
\frac{w_1w_n}{w_{n+1}}  & 0    \\ 0      &  1
\\
\end{array} \hspace{-0.3ex}
\right)
  \left(
\begin{array}{l}
\varphi_{1} \\ \varphi_{0}   \\
\end{array}
\right).
\]
Therefore the
eigenvalue problem (\ref{4}) -- (\ref{5}) is equivalent to the
following one:
\begin{equation}\label{eig}
\left[\frac{1}{w_n}I-B_nS_n(z)\right]\left(
\begin{array}{l}
\varphi_{1} \\ \varphi_{0}   \\
\end{array}
\right)=0.
\end{equation}
Here $I$ is $2\times 2$ identity matrix and
\begin{equation}\label{B_n}
B_n=\diag (\beta_n,1), \hspace{4ex}\hspace{1ex}
\beta_n=\frac{w_{n+1}}{w_1w_n}=e^{ \frac{1}{2}
(\eta_0-\xi_0+\xi_{n}-\eta_{n}) }.
\end{equation}
Thus $z$ is an eigenvalue of $H_n+V_n$ if and only if
\[
\mu^{(1)}_n(z)=\frac{1}{w_n}\hspace{2ex} \hbox{or} \hspace{2ex}
\mu^{(2)}_n(z)=\frac{1}{w_n},
\]
where $ \mu^{(i)}_n(z)$, $i=1,2$ are the eigenvalues of
$B_nS_n(z)$.

Without loss of generality we may suppose that $g\ge 0$. Then, by
the ergodic theorem,
\begin{equation}\label{09}
\lim_{n\to\infty} \hspace{0.5ex}\frac{1}{n}\log
\frac{1}{w_n}\hspace{1ex}=^{^{\hbox{\tiny \hspace{-5ex}Prob.1
}}}\hspace{-0.5ex} g\ge 0.
\end{equation}
Since $\det S_n(z)=c_0/c_n$, $\det B_nS_n(z)$ does not depend on
$z$ and, because of (\ref{B_n})and (\ref{c_}),
\begin{equation}\label{08}
\lim_{n\to\infty} \hspace{0.5ex}\frac{1}{n}\log \det B_nS_n(z)
=\lim_{n\to\infty} \hspace{0.5ex}\frac{1}{n}(\eta_0+\eta_n)
\hspace{1ex}=^{^{\hbox{\tiny \hspace{-5ex}Prob.1
}}}\hspace{-0.5ex}  0.
\end{equation}
Let $r(B_nS_n(z))$ be the spectral radius of $B_nS_n(z)$, i.e.\
\begin{equation}\label{srad}
r(B_nS_n(z))=\max \{|\mu|: \hspace{2ex} \hbox{$\mu$ is an
eigenvalue of $B_nS_n(z)$}\hspace{0.5ex}\}.
\end{equation}
>From (\ref{09}) -- (\ref{08}) we deduce a necessary condition for
$z$ to be an eigenvalue of $H_n +V_n$.  This condition applies to
$P$-almost all $\omega\equiv\{\xi_j, \eta_j, q_j
\}_{j=0}^{\infty}$ and is as follows: If $z$ is an eigenvalue of
$H_n+V_n$ and $n>n(\omega)$ then
\begin{equation}\label{ll1}
\frac{1}{n}\log r(B_nS_n(z))=\frac{1}{n}\log \frac{1}{w_n}.
\end{equation}
It is clear that this condition is necessary but not sufficient.

\medskip

We start with part (b) of Theorem \ref{th1}. Let $K$ be an
arbitrary compact subset of $D_2$. We shall prove that for
$P$-almost all $\omega$ there is an integer $n_0(K,\omega)$ such
that for all $n>n_0(K,\omega)$ equation (\ref{ll1}) cannot be
solved if $z\in K$.

Note that $r (B_nS_n(z)) \le || B_nS_n(z)|| \le ||B_n||
||S_n(z)||$.  Therefore,
\begin{eqnarray}\label{01}
\frac{1}{n}\log r (B_nS_n(z)) &\le& \frac{1}{n}\log ||B_n|| +
\frac{1}{n}\log ||S_n(z)|| \\\label{02}
   &=& o(1) + \frac{1}{n}\log ||S_n(z)||,
   \hspace{4ex} \hbox{for almost all $\omega$}.
\end{eqnarray}

Note that we only have to consider the case when $g>0$. For, if
$g=0$ then  $D_2=\emptyset$ and we have nothing to prove. Obviously,
one can find $\varepsilon >0$ such that
\[
\sup_{z\in K } \bar \gamma (z) \le g - \varepsilon .
\]
But then, by Theorem \ref{Th2.5b} in Appendix, we have, for
$P$-almost all $\omega $,
\[
 \limsup_{n\to\infty} \left\{ \sup_{z\in K} \frac{1}{n} \log
 ||S_n(z)||\right\}
  \le \sup_{z\in K} \bar\gamma (z)
  \le  g -\varepsilon
\]
and hence, by (\ref{01}) -- (\ref{02}),
\[
\limsup_{n\to\infty}\left\{\sup_{z\in K} \frac{1}{n} \log r
(B_nS_n(z))\right\}\le g -\varepsilon .
\]
Thus, for $P$-almost all $\omega$, there exists an integer $n_1(K,
\omega)$ such that for all $n>n_1(K, \omega) $
\begin{equation}\label{04}
\frac{1}{n}\log r (B_nS_n(z))  \le g-\frac{\varepsilon}{2}
\hspace{3ex} \hbox{for all $z\in K$}.
\end{equation}
Relations (\ref{04}) and (\ref{09}) contradict equation
(\ref{ll1}). This proves part (b) of Theorem \ref{th1}.

\bigskip

Now we shall prove part (a) of Theorem \ref{th1}. For this we need
the following general (deterministic) result.

\bigskip

\begin{Lemma}
\label{l1}  Let $q_1, \ldots , q_n$ be real and
$c_0,\ldots, c_n$ be positive.
For any two complex numbers $\phi_0$ and $\phi_1$
define recursively a sequence of complex numbers $\phi_2, \ldots ,
\phi_{n+1}$ as follows:
\begin{equation}\label{rrr}
c_{k}\phi_{k+1}=(q_k-z)\phi_k - c_{k-1}\phi_{k-1}, \;\;\; k=1,2,\ldots,n.
\end{equation}
Denote by $S_n$ the $2\times 2$ matrix which maps
$(\phi_1,\phi_0)^T$ into $(\phi_{n+1},\phi_n)^T$ obtained as
prescribed above, i.e.\  $S_n(z)=A_n\cdot \ldots \cdot A_1$, where
$A_k$ is given by (\ref{10}). Then for every $z\in \C_{+}$ the
matrix $S_n$ has two linearly independent eigenvectors $(u^*,1)^T$
and $(v^*,1)^T$ such that
\begin{equation}\label{km0}
\im u^* \le -\frac{\im z}{c_n},\;\;\;\; |u^*|\le
\frac{|q_n-z|}{c_n} +\frac{c_{n-1}^2}{c_n\im z}
\end{equation}
and
\begin{equation} \label{kp01}
\im v^*\ge 0, \;\;\;\; |v^*|\le \frac{c_{0}}{\im z}
\end{equation}
\end{Lemma}

\smallskip

\noindent \emph{Proof}. Define $u_1, \ldots, u_{n+1}$ as follows:
$ u_1=u$ and $u_{k+1}=f_k(u_k)$, $ k=1,\ldots n$, where
\[
f_k(u)=\frac{q_k-z}{c_k} -\frac{c_{k-1}}{c_k} \frac{1}{u}.
\]
Obviously,
$
u_{n+1}= F_n(u)$ where $ F_n(u)=f_n(f_{n-1}(\ldots f_1 (u))\ldots
)$.

Since $\im z $ and all $c_k$ are positive, each of the functions
$f_k$ maps $\C_{-}$ into itself and
\[
\im f_n(u)\le -\frac{\im z}{c_n}, \hspace{3ex} |f_n(u)|\le
\frac{|q_n-z_n|}{c_n}+\frac{c_{n-1}}{c_n}\frac{1}{|u|}
\hspace{3ex} \forall u\in \C_{-}.
\]
Therefore, $F_n$ maps continuously the compact set $Q$ defined by
the inequalities (\ref{km0}) into itself. By the fixed point
theorem, there exists $u^*$ in $Q$ such that $u^{*}=F_n (u^{*})$,
i.e.\ if $u_{n+1}=u^*$ given that $u_1=u^*$.

Now set $\varphi_0=1$, $\varphi_1=u^{*}$ and iterate (\ref{rrr})
to obtain $\varphi_k$, $k=2, \ldots n+1$ using this initial data.
Since $ \frac{\varphi_{1}}{\varphi_{0}}=u^{*}$ and
\[
\frac{\varphi_{k+1}}{\varphi_{k}}=\frac{(q_k-z)}{c_k} -
\frac{c_{k-1}}{c_k} \frac{\varphi_{k-1}}{\varphi_{k}},\;\;\;\;
k=1,2, \ldots, n,
\]
we have that $\frac{\varphi_{k+1}}{\varphi_{k}}=u_{k+1}$ for all
$k$ with the $u_k$ as above. In particular,
$\frac{\varphi_{n+1}}{\varphi_{n}} = u_{n+1}=u^{*}$ and hence
$\varphi_{n+1}=\varphi_nu^{*}$. But then
\begin{equation}
\label{v1} S_n \left(
\begin{array}{l}
u^{*} \\ 1
\end{array}
\right) = \varphi_n  \left(
\begin{array}{l}
u^{*} \\  1
\end{array}
\right)
\end{equation}
and $(u^{*}, 1)^T$ is an eigenvector of $S_n(z)$ with $u^{*}$
satisfying (\ref{km0}).

To construct the other eigenvector one can iterate recursion
(\ref{rrr}) in the opposite direction. Namely, set
$\varphi_{n+1}=v_{n+1}$, $\varphi_n=1$ and use (\ref{rrr})
backwards to obtain the remaining $\varphi_k$. Similarly to what
we have done before, write (\ref{rrr}) as
$
v_k=\tilde f_k( v_{k+1})$, $k=n, \ldots 1$, where now
$v_k=\frac{\varphi_{k}}{\varphi_{k-1}}$ and
\[
\tilde f_k(v)=\frac{c_{k-1}}{q_k-z-c_k v}.
\]
Then $v_1= \tilde F_n(v_{n+1})$, where $\tilde F_n$ is the
composition of $\tilde f_1, \tilde f_2, \ldots, \tilde f_{n}$.
Each of these functions is continuous in the closure of the upper
half of the complex plane and maps this set into itself. Since
\[
|\tilde f_1(u)| \le \frac{c_{0}}{\im z}  \hspace{4ex} \forall u
\in \C_{+},
\]
$\tilde F_n$ maps the compact set $\tilde Q$ defined by the
inequalities in (\ref{kp01}) into itself. By the fixed point
theorem, there exists $v^* \in \tilde Q$ such that $\tilde
F_n(v^*)=v^*$, i.e. $\varphi_1=v^{*}\varphi_0$ given that
$\varphi_{n+1}=v^{*}$ and $\varphi_n=1$. But then
\begin{equation}
\label{w} S_n^{-1}(z) \left(
\begin{array}{l}
v^{*} \\ 1
\end{array}
\right) = \varphi_0 \left(
\begin{array}{l}
v^{*} \\ 1
\end{array}
\right)
\end{equation}
and $(v^{*}, 1)^T$ is an eigenvector of $S_n(z)$ with $v^{*}$
satisfying (\ref{kp01}).

It is apparent that the two constructed eigenvectors are linearly
independent. \hfill $\blacksquare$

\bigskip

\begin{Lemma}\label{l1a}
Let $S_n(z)$ be as in Lemma \ref{l1} and $B_n=\diag (\beta_n, 1)$
with $\beta_n >0$. Then for every $z\in \C_{+}$ the matrix
$B_nS_n$ has two linearly independent eigenvectors $(u_n,1)^T$ and
$(v_n,1)^T$ such that
\begin{equation}
\label{km} \im u_n \le -\frac{\beta_n \im z }{c_n},\;\;\;\;
|u_n|\le \frac{\beta_n|q_n-z|}{c_n} +\frac{\beta_n
c_{n-1}^2}{c_n\im z}
\end{equation}
and
\begin{equation}
\label{kp} \im v_n\ge 0, \;\;\;\; |v_n|\le \frac{c_{0}}{\im z}
\end{equation}
\end{Lemma}

\bigskip

\emph{Proof}. $B_nS_n(z)=\tilde A_n \cdot A_{n-1} \cdot \ldots
\cdot A_1$, where $A_1, \ldots , A_{n-1}$ as before (see
(\ref{10}) -- (\ref{tranmat})) and
\[
\tilde A_n =\left(
\begin{array}{l l}
\frac{(q_{n}-z)\beta_n}{c_n}  & -\frac{c_{n-1}\beta_n}{c_n}    \\
1 & \phantom{-} 0
\\
\end{array}
\right)
\]
Since $\beta_n>0$, Lemma \ref{l1} applies. \hfill $\blacksquare$

\bigskip

We now return to our eigenvalue problem  (\ref{4}) -- (\ref{5})
and to the matrices $S_n(z)$ and $B_n$  associated with this
problem, i.e. now $c_k$ are given by (\ref{c_}) and $\beta_n$ by
(\ref{B_n}). Set
\begin{equation} \label{T}
T_n= \left(
\begin{array}{l l}
u_n & v_n \\ 1 & 1
\end{array}
\right)
\end{equation}
where $(u_n, 1)^T$ and $(v_n, 1)^T$ are the eigenvectors of
$B_nS_n(z)$ obtained in Lemma \ref{l1a}.

\bigskip

\begin{Lemma}
\label{l3} Assume (\ref{ba1}) -- (\ref{ba2}). Then for $P$-almost
all $\{\xi_k, \eta_k, q_k \}_{k=0}^{\infty}$ the following
statement holds: For all $z\in \C_{+}$
\begin{equation}\label{kappa1}
\lim_{n\to \infty } \frac{1}{n} \log || T_n|| ||T_n^{-1}|| =0.
\end{equation}
The convergence in (\ref{kappa1}) is uniform in $z$ on every compact set
in $\C_{+}$.
\end{Lemma}
\smallskip

\noindent \emph{Proof}. For any stationary ergodic sequence of
random variables $X_n$ with finite first moment $\lim_{n\to\infty}
X_n/n = 0$ with probability 1. By (\ref{norm}),
\[
0\le \frac{1}{n}\log ||T_n|| ||T_n^{-1}||\le \frac{1}{n}\log
\frac{(|u_n|+|v_n|+2)^2}{|u_n-v_n|}.
\]
To complete the proof, apply inequalities (\ref{km}) --
(\ref{kp}). \hfill $\blacksquare$

\bigskip

Recall that $r(B_nS_n(z))$ is used to denote the spectral radius
(\ref{srad}).
\bigskip

\begin{Lemma}
\label{l4} Assume (\ref{ba1}) -- (\ref{ba2}). Then for $P$-almost
all $\{\xi_k, \eta_k, q_k \}_{k=0}^{\infty}$ the following
statement holds: For all $z\in \C_{+}$
\begin{equation}\label{lambda}
\lim_{n \to \infty} \frac{1}{n}\log r (B_nS_n(z))= \bar \gamma
(z),
\end{equation}
where $\bar \gamma (z)$ is the Lyapunov exponent (\ref{13}). The
convergence in (\ref{kappa1}) is uniform on every compact set in
$\C_{+}$.
\end{Lemma}

\smallskip

\noindent \emph{Proof}. It follows from Lemma \ref{l1a} that
$
B_nS_n(z) =T_n\Lambda_n T_n^{-1},
$
where $\Lambda_n $ is the diagonal matrix of eigenvalues of
$B_nS_n(z)$ corresponding to $T_n$. Then
\[
\frac{1}{||T_n||||T_n^{-1}||}\le \frac{||B_nS_n||}{||\Lambda_n
||}\le ||T_n||||T_n^{-1}||.
\]
With our choice (\ref{norm}) of the matrix norm, $||\Lambda || = r
(B_nS_n(z))$ and, by Lemma \ref{l3}, for $P$-almost all $\{\xi_k,
\eta_k, q_k \}_{k=0}^{\infty}$,
\[
\lim_{n \to \infty} \frac{1}{n}\log \frac{|| B_nS_n(z)||}{ r
(B_nS_n(z))}=0 \hspace{3ex} \hbox{uniformly in $z$ on compact
subsets of $\C_{+}$.}
\]
On the other hand, for $P$-almost all $\{\xi_k, \eta_k, q_k
\}_{k=0}^{\infty}$,
\[
\lim_{n \to \infty} \frac{1}{n}
\log\frac{||B_nS_n(z)||}{||S_n(z)||}=0 \hspace{3ex}
\hbox{uniformly in $z$.}
\]
This follows from the obvious inequalities
\[
\frac{||S_n(z)||}{||B^{-1}_n||}\le || B_nS_n(z)||\le
||B_n||||S_n(z)||.
\]
Now the statement of Lemma follows from Theorem \ref{Th2.5} of
Appendix.  \hfill $\blacksquare$

\bigskip

With Lemma \ref{l4} in hand, we are a in a position to prove part
(a) of Theorem \ref{th1}. Let $K$ be a compact subset of
$D_1\backslash \R$. As  $\bar \gamma (z)$ is continuous in $K$ and
$\bar \gamma (z) > g$ there, one can find an $\varepsilon >0$ such
that
\[
\min_{z\in K} \bar \gamma (z) \ge  g +\varepsilon.
\]
>From this, by Lemma \ref{l4}, for almost all $\omega = \{\xi_k,
\eta_k, q_k \}_{k=0}^{\infty}$, there exists an integer $n_1(K,
\omega)$ such that for all $n>n_1(K, \omega)$
\begin{equation}\label{66}
\frac{1}{n}\log r(B_nS_n(z)) \ge g+\frac{\varepsilon}{2}
\hspace{3ex} \forall z\in K.
\end{equation}
In view of (\ref{09}), (\ref{66}) contradicts (\ref{ll1}). Theorem
\ref{th1}  is proved.

\section{Distribution of eigenvalues}
\label{section3}

\bigskip

We need to introduce more notations and to recall few
elementary facts from potential theory.

Let $M_n$ be an $n\times n$ matrix.  We denote by
$d\nu_{\!\scriptscriptstyle{M_n}}$ the measure on $\C$ that assigns to each
of the $n$ eigenvalues of $M_n$ the mass $\frac{1}{n}$. This
measure describes the distribution of eigenvalues of $M_n$ in the
complex plane in the following sense. For any rectangle $K \subset
\C$
\[
\nu (K; M_n)= \int_K d\nu_{\!\scriptscriptstyle{M_n}}
\]
gives the proportion of the eigenvalues of $M_n$ that are in $K$.
The eigenvalues are counted according to their multiplicity.

The measure $d\nu_{\!\scriptscriptstyle{M_n}}$ can be obtained
from the characteristic polynomial of $M_n$ as follows. Let
\begin{eqnarray}\label{p1}
p(z;M_n)&=&\frac{1}{n}\log |\det (M_n-zI_n)|
 \\ &=& \label{p2} \int_{\C} \log
 |z-\zeta| d\nu_{\!\scriptscriptstyle{M_n}}(\zeta)
\end{eqnarray}
In view of (\ref{p2}), $p(z;M_n)$ is the potential of the
eigenvalue distribution of $M_n$. Obviously, $p(z;M_n)$ is locally
integrable in $z$. Then for any sufficiently smooth function
$f(z)$ with compact support
\begin{eqnarray*}
\int_{\C}\log |z-\zeta|\Delta f(z) d^2z
&=&\lim_{\varepsilon\downarrow 0}\int_{|z-\zeta|\ge \varepsilon }
\log |z-\zeta|\Delta f(z) d^2z\\
  &=& 2\pi f(\zeta ),
\end{eqnarray*}
by Green's formula. Hence
\begin{equation}\label{78}
 \frac{1}{2\pi} \int_{\C}
p(z; M_n)\Delta f(z) d^2z =
\int_{\C}f(z)d\nu_{\!\scriptscriptstyle{M_n}}\!(z).
\end{equation}
Here $\Delta$ is the two-dimensional Laplacian and $d^2z$ is the
element of area in the complex plane. Both $p(z;M_n)$ and
$d\nu_{\!\scriptscriptstyle{M_n}}$ define distributions in the
sense of the theory of distributions and Eq.\ (\ref{78}) can be
also read as the equality
$d\nu_{\!\scriptscriptstyle{M_n}}\!(z)=\frac{1}{2\pi}\Delta
p(z;M_n)$ where now $\Delta$ is the distributional Laplacian. More
generally, it is proved in potential theory that, under
appropriate conditions on $d\nu $, $ d\nu (z)
=\frac{1}{2\pi}\Delta p(z)$, where $p(z)=\int \log |z-\zeta|
d\nu_{\!\scriptscriptstyle{M_n}}\!(\zeta)$ is the potential of
$d\nu$. This Poisson's equation relates measures and their
potentials.

In this section we shall calculate the limit of
$d\nu_{\!\scriptscriptstyle{J_n}}$ for matrices (\ref{3}),
obtaining the potential of the limiting measure in terms of the
integrated density of states $N(\lambda)$ (see (\ref{N})). Our
calculation makes use of relation (\ref{simtr}) according to which
the asymmetric matrix $J_n$ is, modulo a similarity
transformation, a rank 2 perturbation of the symmetric matrix
$H_n$ (see (\ref{7})). The low rank of the perturbation allows to obtain
explicit formulas describing the change in
location of the eigenvalues. In our case,
\begin{equation}\label{det}
\det (J_n-zI_n)=\det(H_n+V_n-zI_n)= d(z;\, H_n,V_n)\det
(H_n-zI_n),
\end{equation}
were
\begin{eqnarray}\label{dn}
d(z;\, H_n,V_n)&=& \det [I_n+V_n(H_n-zI_n)^{-1}] \\
            &=& \label{d_n}
(1+a_nG_{n1})(1 + b_nG_{1n})-a_nb_nG_{11}G_{nn}
\end{eqnarray}
with $a_n$ and $b_n$ being the top-right and left-bottom corner
entries of $V_n$ and $G_{lm}$ standing for the $(l,m)$ entry of
$(H_n-zI_n)^{-1}$.

One can easily obtain (\ref{d_n}) from (\ref{dn}) with the help of
a little trick. Write $V_n$ in the form $V_n=A^TB$, where $A$ and
$B$ are the following $2\times n$ matrices:
\[
A= \left(
\begin{array}{c c c c c}
a_n  & 0   & \ldots  & 0   & 0
\\
0  & 0   & \ldots  & 0   & 1
\\
\end{array}
\right)\ \;\;\;\;\; B= \left(
\begin{array}{c c c c c}
0 & 0   & \ldots  & 0   & 1
\\
b_n  & 0   & \ldots  & 0   & 0
\\
\end{array}
\right).
\]
Then the $n\times n$ determinant in (\ref{dn}) reduces to a
$2\times 2$ determinant, as $\det [I_n+A^TB(H_n-zI_n)^{-1}] = \det
[I_2+B(H_n-zI_n)^{-1}A^T] $ and the latter, being expanded, gives
(\ref{d_n}).

Eqs.\ (\ref{det}) -- (\ref{d_n}) yield the following relationship
between the potentials of $d\nu_{\!\scriptscriptstyle{J_n}}$ and
$d\nu_{\!\scriptscriptstyle{H_n}}$:
\begin{eqnarray}
p(z;J_n)&=& p(z;H_n+V_n)\nonumber \\
        &=& p(z;H_n)+ \frac{1}{n}\log |d (z; H_n,
V_n)|. \label{J2H}
\end{eqnarray}
The measures $d\nu_{\!\scriptscriptstyle{H_n}}$ are all supported
on the real axis where they converge to the measure $dN(\lambda )$
when $n\to\infty$. This implies the convergence of their
potentials to
$
\Phi (z)= \int_{\R} \log |z-\lambda |dN(\lambda)
$
for all non-real $z$, see Theorem \ref{Th2.3} in Appendix.
$\Phi(z)$ is the potential of the limiting distribution of
eigenvalues for $H_n$. Thus the main part of our calculation of
$p(z;H_n)$ in the limit $n\to\infty$ is evaluating the
contribution of $\frac{1}{n}\log |d (z; H_n, V_n)|$ to $p(z;J_n)$.
We can do this for all non-real $z$  lying off the curve ${\cal
L}$ (see (\ref{14}) and (\ref{0018})). The corresponding result is central to our
considerations and we state it as Theorem \ref{sec3:th1} below.
Note that, by the Thouless formula (\ref{18}), $\bar \gamma (z)=|g|$
is equivalent to $\Phi (z) = \max (\E\xi_0,\E \eta_0)$ and thus
(cf.\  (\ref{14a}))
\begin{equation} \label{d_1}
D_1=\{z: \hspace{1ex}\Phi (z)>\max (\E\xi_0, \E\eta_0)\ \}
\hspace{3ex} \hbox{and} \hspace{3ex} D_2=\{z:\hspace{1ex} \Phi
(z)< \max (\E\xi_0, \E\eta_0)\ \}.
\end{equation}

\bigskip

\begin{Th}\label{sec3:th1}
Assume (\ref{ba1})--(\ref{ba2}). Then for $P$-almost all $\{\xi_j, \eta_j, q_j
\}_{j=0}^{\infty}$,
\begin{eqnarray}\label{F1}
\lim_{n\to\infty} p(z;J_n) &=&  \Phi (z) \hspace{14ex} \forall
z\in D_1\backslash\R;
\\ \lim_{n\to\infty} p(z;J_n)
&=& \label{F2} \max (\E\xi_0, \E\eta_0)  \hspace{3ex}  \forall
z\in D_2\backslash\R.
\end{eqnarray}
The convergence in (\ref{F1}) -- (\ref{F2}) is uniform in $z$ on
every compact set in $D_1\backslash\R$ and $D_2\backslash\R$
respectively.
\end{Th}

\medskip

\noindent \emph{Proof}. In view of Eq.\ (\ref{J2H}) and Theorem
\ref{Th2.3} in Appendix, we only have to prove that, with
probability one, for any compact sets $K_1\subset D_1\backslash
\R$ and $K_2\subset D_2\backslash \R$
\begin{equation}\label{lim1}
\lim_{n\to\infty} \frac{1}{n} \log |d(z; H_n,V_n)| =  0
\hspace{24.7ex} (\hbox{uniformly in}\ z\in K_1\subset
D_1\backslash \R),
\end{equation}
\vspace{-4ex}
\begin{equation}\label{lim2}
\lim_{n\to\infty} \frac{1}{n} \log |d(z; H_n,V_n)| = \max
(\E\xi_0, \E\eta_0) - \Phi (z)  \hspace{3ex} (\hbox{uniformly in}\
z\in K_2\subset D_2\backslash \R).
\end{equation}
Let us write $d(z;H_n,V_n)$ in the form
\begin{equation}\label{99}
d(z;H_n,V_n) = a_nG_{n1} + b_nG_{1n} +a_nb_nG_{n1}G_{1n}
+(1-a_nb_nG_{11}G_{nn})
\end{equation}
and estimate the four terms on the right-hand side (r.h.s.) of
(\ref{99}). Recall that $a_n$ and $b_n$ are the corner entries of
$V_n$ and the $G$'s  are the corner
entries of $(H_n-zI_n)^{-1}$. In particular,
\[
G_{1n}=G_{n1}=\frac{\prod_{j=1}^{n-1}c_j}{\det
(H_n-zI_n)}=\frac{e^{\frac{1}{2}\sum_{j=1}^{n-1} (\xi_j+\eta_j)
}}{\det (H_n-zI_n)}.
\]
Under assumptions (\ref{ba1}) -- (\ref{ba2}), on a set of full
probability measure,
\begin{equation}\label{90}
a_n=-e^{n [\frac{1}{2}\E(\xi_0-\eta_0)+o(1)]},\hspace{3ex}
b_n=-e^{-n[\frac{1}{2}\E(\xi_0-\eta_0)+o(1)]},
\end{equation}
when $n\to\infty$ and
\[
e^{\frac{1}{2}\sum_{j=1}^{n-1} (\xi_j+\eta_j)
}=e^{n[\frac{1}{2}\E(\xi_0+\eta_0)+o(1)]}.
\]
On the other hand,
\[
|\det (H_n-zI_n) |=e^{np(z;H_n)}=e^{n[\Phi (z)-r(z;H_n)]},
\]
where $r(z;H_n)= p(z;H_n)-\Phi(z)$. According to Theorem
\ref{Th2.3} in Appendix for  $P$-almost all $\{\xi_j, \eta_j, q_j
\}_{j=0}^{\infty}$
\begin{equation}\label{100}
\lim_{n\to\infty} r(z; H_n)=0 \hspace{0.5cm} (\hbox{uniformly in
$z$ on every compact set $K\subset \C\backslash \R$}).
\end{equation}
Therefore, with probability one,
\begin{equation}\label{101}
|a_nG_{n1}|=e^{n[\E\xi_0-\Phi(z)+o_{z}(1)]} \hspace{2ex}
\hbox{and} \hspace{2ex}
|b_nG_{1n}|=e^{n[\E\eta_0-\Phi(z)+o_{z}(1)]},
\end{equation}
where the $o_z(1)$ terms vanish when $n\to\infty$
uniformly in $z$ on every compact set  $K\subset \C\backslash \R$.

To estimate the third term in the r.h.s.\ of (\ref{99}), recall
the Thouless formula (\ref{18}). As the Lyapunov exponent $\bar
\gamma (z)$ is non-negative everywhere in the complex plane and
$\Phi (z)$, for every fixed $\re z$, is an increasing function of
$|\im z|$, we have that
\begin{equation}\label{97}
\Phi (z) > \frac{1}{2}\E (\xi_0+\eta_0) \hspace{4ex} \forall z\in
\C\backslash \R.
\end{equation}
Moreover, since $\Phi (z) $ is continuous in $z$ off the
real axis,
\[
M(K):=\min_{z\in K} [\Phi (z) - \frac{1}{2}\E (\xi_0+\eta_0)] >0
\]
for any compact set $K\subset \C\backslash \R$. From this and
(\ref{101}),
\begin{eqnarray} \label{92}
|a_nb_nG_{1n}G_{n1}| &=& e^{n[E(\xi_0+\eta_0)-2\Phi(z)+o_z(1) ]}\\
                     &\le& e^{-n[2 M(K)+o_z(1)]}
                     \hspace{3ex}\forall z\in K. \label{93}
\end{eqnarray}
In other words, the third term on the r.h.s.\ in (\ref{99})
vanishes exponentially fast (and uniformly in $z$ on every compact
set in $\C\backslash \R$) in the limit $n\to\infty$.

The fourth term in (\ref{99}) cannot grow exponentially fast with
$n$. Nor it can vanish exponentially fast. Estimating it from
above is simple. Since $|G_{jj}|\le \frac{1}{|\im
z|}$, $j=1,n$, and $a_nb_n=e^{o(n)}$, we have,
for $P$-almost all $\{\xi_j, \eta_j, q_j
\}_{j=0}^{\infty}$, that
\begin{equation}\label{91}
|1-a_nb_nG_{11}G_{nn}|\le 1+\frac{e^{o(n)}}{|\im z|^2}
\hspace{3ex}\forall z\in \C\backslash\R
\end{equation}
with the $o(n)$ term being independent of $z$.

Estimating
$|1-a_nb_nG_{11}G_{nn}|$ from below is less trivial. We do this
with the help of the two Propositions stated below. The first one
is elementary and the second one is a standard result from
spectral theory of random operators \cite{CL}.

\bigskip

\begin{Prop}\label{prop1}
Let $C_{\alpha}=\{z\in \C: \hspace{2ex}\alpha \le \arg z \le
\alpha +\pi \hspace{1ex} \}$, where $0\le \alpha \le\pi$. Then
\[
\min_{z\in C_{\alpha}} |1-z|=\sin \alpha
\]
\end{Prop}

\medskip

\begin{Prop}\label{prop2}
Under assumptions (\ref{ba1}) -- (\ref{ba2}), there exists a set
$\Omega_0\subset \Omega$ of full probability measure such that for
every $\omega=\{q_k, \xi_k,\eta_k \}_{k=0}^{\infty} \in \Omega_0$
\begin{equation}\label{pr2}
\lim_{n\to\infty}
\left[(H_n-zI_n)^{-1}\right]_{11}=\int_{-\infty}^{+\infty}\frac{d\sigma
(\lambda; \omega) }{\lambda -z}  \hspace{4ex} \forall z\in
\C\backslash\R,
\end{equation}
where $\sigma (\lambda; \omega) $, for every $\omega \in \Omega_0
$, is a measure on $\R$ with unit total mass. The convergence in
(\ref{pr2}) is uniform on every compact set in $\C\backslash\R$.
\end{Prop}

\medskip

\noindent {\bf Remark}. For almost all realizations $\omega$, the
semi-infinite matrix
\[
\left(
\begin{array}{c c c c}
q_{1}  & -c_{1}   &            &
\\
-c_{1} &  q_2  & -c_{2}         &
\\
       &\hspace{-2ex} \ddots    &  \hspace{-2ex} \ddots   & \hspace{-2ex}
       \ddots
\end{array}
\right),
\]
where $c_k=e^{\frac{1}{2}(\xi_k+\eta_k)}$, specifies a selfadjoint
operator $H_{+}(\omega)$ on $l^2(\Z_{+})$. Proposition \ref{prop2}
is a consequence of the selfadjointness of $H_{+}(\omega)$, and
the measure $\sigma (\lambda; \omega)$ is simply the
(1,1) entry of the resolution of identity for $H_{+}(\omega)$.

\bigskip

Let us set
\[
\alpha (z;H_n)= \left\{
\begin{array}{l l}
\arg \big[(H_n-zI_n)^{-1}\big]_{11}, &\hspace{3ex} \hbox{if}\ \im
z
>0\\[1ex] \arg \big[(zI_n-H_n)^{-1}\big]_{11}, & \hspace{3ex} \hbox{if}\ \im z
<0.
\end{array}
\right.
\]
$G_{11}=\big[(H_n-zI_n)^{-1}\big]_{11}$, as a function of $z$,  maps
the upper (lower) half of the complex plane into itself, and so
does $G_{nn}$. Therefore,
\[
0<\alpha(z;H_n)<\pi  \hspace{3ex} \forall z\in \C\backslash \R
\]
and
\[
\alpha(z;H_n) \le \arg(a_nb_nG_{11}G_{nn})\le \alpha(z;H_n) +\pi
\hspace{3ex} \forall z\in \C\backslash \R.
\]
Then by Proposition \ref{prop1}
\[
|1-a_nb_nG_{11}G_{nn}|\ge \sin\alpha(z;H_n) \hspace{3ex} \forall
z\in \C\backslash \R.
\]
Obviously, $\sin \alpha(z;H_n)=|\im G_{11}|/|G_{11}|$ and by
Proposition \ref{prop2} for every $\omega\in \Omega_0 $ and every
compact set $K\subset \C\backslash \R$
\begin{equation}\label{111}
\lim_{n\to\infty} \sin\alpha(z;H_n) = \frac{\displaystyle{
\int_{-\infty}^{+\infty}\frac{ |\im z|\ d\sigma (\lambda; \omega)
}{|\lambda
-z|^2}}}{\displaystyle{\left|\int_{-\infty}^{+\infty}\frac{d\sigma
(\lambda; \omega) }{\lambda -z}\right|}} \hspace{6ex}
(\hbox{uniformly in}\ z\in K).
\end{equation}
The r.h.s.\ in (\ref{111}) is positive and continuous in $z\in K$.
Therefore it is bounded away from zero uniformly in $z\in K$.
Hence, for every $\omega \in \Omega_0$ (set of full probability
measure) and for every compact set $K\subset \C\backslash \R$
\begin{equation}\label{112}
\liminf_{n\to\infty} |1-a_nb_nG_{11}G_{nn}| \ge C(K;\omega)> 0
\hspace{3ex} (\hbox{uniformly in}\ z\in K),
\end{equation}
where the constant $C(K;\omega)$ depends only on $K$ and $\omega$.

Now we are in a position to prove (\ref{lim1}) and (\ref{lim2}).
Let $z\in K\subset D_1\backslash\R$. Then
\[
\Phi (z) > E\xi_0 \hspace{3ex} \hbox{and} \hspace{3ex} \Phi (z) >
E\eta_0 \hspace{3ex} \forall z\in K.
\]
Therefore, by (\ref{101}), the first three terms on the r.h.s.\ in
(\ref{99}) vanish exponentially fast (and uniformly in $z\in K$)
in the limit $n\to\infty$. But then, with probability one,
\[
\limsup_{n\to\infty} \frac{1}{n}\log |d(z; H_n, V_n)|\le 0
\hspace{3ex} (\hbox{uniformly in $z\in K$}),
\]
in view of (\ref{91}), and
\[
\liminf_{n\to\infty} \frac{1}{n}\log |d(z; H_n, V_n)|\ge 0
\hspace{3ex} (\hbox{uniformly in $z\in K$}),
\]
in view of (\ref{112}). This proves (\ref{lim1}).

Let $K$ be any compact set in $D_2\backslash\R$. Then, by the
definition of $D_2$ and (\ref{97}),
\[
\min (\E\xi_0, \E\eta_0) < \Phi (z) < \max (\E\xi_0, \E\eta_0).
\]
At this point we may assume, without loss of generality, that
$\E\xi_0\not=\E\eta_0$. For, if $\E\xi_0=\E\eta_0$ then
$D_2\backslash \R$ is empty. This follows from (\ref{97}).

Now, if $\E\xi_0>\E\eta_0$ then $\E\eta_0 < \Phi (z) < \E\xi_0$
for all $z\in K$ and, by (\ref{101}), (\ref{92}) -- (\ref{91}),
the first term on the r.h.s.\ in (\ref{99}) dominates the other
terms.  In this case, with probability one,
\[
\lim_{n\to\infty} \frac{1}{n} \log |d(z; H_n,V_n)|  =  \E\xi_0 -
\Phi (z) \hspace{2ex} (\hbox{uniformly in}\ z\in K).
\]
Similarly, if $\E\xi_0 < \E\eta_0$, then it is the second term
that dominates and, with probability one,
\[
\lim_{n\to\infty} \frac{1}{n} \log |d(z; H_n,V_n)|  =  \E\eta_0 -
\Phi (z) \hspace{2ex} (\hbox{uniformly in}\ z\in K).
\]
Theorem \ref{sec3:th1} is proved. \hfill $\blacksquare$

\bigskip

We shall now deduce from Theorem \ref{sec3:th1} the weak
convergence of the eigenvalue distributions
$d\nu_{\!\scriptscriptstyle{J_n}}$ to a limiting measure in the
limit $n\to\infty$. In doing this we shall follow Widom
\cite{Widom1,Widom2}  who proved that the almost everywhere
convergence of the potentials $p(z; M_n)$ (\ref{p2}) of atomic
measures $d\nu_{\!\scriptscriptstyle{M_n}}$ implies the weak
convergence of the measures themselves provided they are supported
inside a \emph{bounded} domain in the complex plane. Under
assumptions (\ref{ba1}) -- (\ref{ba2}), the spectra of $J_n$
(\ref{3}) are not necessarily confined to a bounded domain. To
extend Widom's argument to our case we estimate the contribution
of the tails of $d\nu_{\!\scriptscriptstyle{J_n}}$  to the
corresponding potentials.

Define
\begin{equation}\label{p}
p(z)=\max [\Phi (z), \E\xi_0, \E\eta_0 ].
\end{equation}
This function coincides with the r.h.s.\ of (\ref{F1}) -
(\ref{F2}) and is continuous everywhere in the complex plane
except may be the set $\Sigma $, see (\ref{Sigma}).

$\Phi (z)$ is a subharmonic function \cite{CS} and so is $p(z)$. For, the
maximum of two subharmonic functions is subharmonic too.
Therefore, $\Delta p(z)$ is non-negative in the sense of
distribution theory and $\frac{1}{2\pi} \Delta p(z)$ defines a
measure in $\C$ which we denote by $d\nu$,
\begin{equation}\label{nu}
d\nu (z)=\frac{1}{2\pi}\Delta p(z).
\end{equation}
First, we prove that in the limit $n\to\infty$ the potentials
$p(z;J_n)$ converge to $p(z)$, for $P$-almost all
$\{\xi_k, \eta_k, q_k \}_{k=0}^{\infty}$, in the sense of
distribution theory.

In the Lemma below $C_0(\C)$ is the space of continuous on $\C$
functions with compact support, and
$C_0^{\infty}(\C)$ is the subspace of $C_0(\C)$ of those functions which are
infinitely differentiable in $\re z$ and $\im z$.

\bigskip

\begin{Lemma}\label{sec3:th2}
Assume (\ref{ba1}) -- (\ref{ba2}). Then on a set of full
probability measure, for every $f\in C_0(\C)$, and in particular
for every  $f\in C_0^{\infty} (\C)$,
\begin{equation}\label{t0}
\lim_{n\to\infty}\int_{\C} f(z) p(z;J_n) d^2z = \int_{\C} f(z)
p(z) d^2z.
\end{equation}
\end{Lemma}

\medskip

\noindent \emph{Proof}. Since $p(z)$ is subharmonic, $p(z)\in
L^1_{\hbox{\tiny loc}}(\C)$, so the integral on the r.h.s.\ of
(\ref{t0}) makes sense. Let $L_{\delta}=\{z\in \C: \hspace{2ex}
\dist (z, \R\cup {\cal L}) \le \delta \}$. By Theorem
\ref{sec3:th1}, on a set of full probability measure $\Omega_0$,
\[
\lim_{n\to\infty} p(z;J_n) = p(z) \hspace{3ex} \hbox{uniformly in
$z$ on compact subsets of $\C\backslash (\R\cup {\cal L})$}.
\]
It follows from this that on the same set $\Omega_0$
\begin{equation}\label{t1}
\lim_{n\to\infty} \int_{\C\backslash L_{\delta}} f(z)p(z;
J_n)d^2z= \lim_{n\to\infty} \int_{\C\backslash L_{\delta}}
f(z)p(z)d^2z
\end{equation}
for every continuous $f(z)$ with compact support and for every
$\delta > 0$.

Since $f(z)$ has compact support, $f(z)p(z)\in L^1(\C)$. Therefore,
\begin{equation}\label{t2}
\forall \varepsilon
> 0\hspace{2ex} \exists \delta >0: \hspace{6ex} \int_{L_{\delta}}
|f(z)p(z)|d^2z < \varepsilon
\end{equation}
and to complete the proof we only need to show that the same is
true for $f(z)p(z;J_n)$, uniformly in $n$. More precisely, it will
suffice to prove the following statement. On a set of full
probability measure, for every continuous function $f$ with
compact support
\begin{equation}\label{eps}
\forall \varepsilon >0 \hspace{2ex} \exists \delta >0:
\hspace{3ex} \limsup_{n\to\infty} \int_{L_{\delta}} |f(z)p(z;J_n)|
d^2z < \varepsilon.
\end{equation}
Obviously, (\ref{t1}) together with (\ref{t2}) -- (\ref{eps})
imply (\ref{t0}).

To prove (\ref{eps}), we break up $p(z;J_n)$ into two parts:
\begin{eqnarray}\label{b1}
p(z;J_n)&=&\frac{1}{n}\sum_{j=1}^n \log |z_j-z|\\
        &=& \label{b2} \frac{1}{n}\sum_{|z_j|\le R} \log |z_j-z| +
        \frac{1}{n}\sum_{|z_j|> R}\log |z_j-z|,
\end{eqnarray}
where $z_1, \ldots , z_n$ are the eigenvalues of $J_n$ and the
summation in the two sums in (\ref{b2}) is over all eigenvalues of
$J_n$ satisfying the inequalities $|z_j|\le  R$ and $|z_j|> R$
respectively. The first term in (\ref{b2}) is bounded from above
but is unbounded from below due to the log-singularities at $z_j$.
On the contrary, the second term is bounded from below, provided $z$
is separated from the boundary of the disk $|z|\le R$, but may be
unbounded from above due to large values of $|z_j-z|$. We shall
treat these two terms separately.

The required  estimate on the integral in (\ref{eps}) involving
the first term in the break-up of $p(z;J_n)$ (\ref{b2}) can be
obtained using the property of local integrability of $\log |z|$.
Recall that a family of functions $\{h_{\alpha}(z) \}_{\alpha \in
{\cal A} }$ is called uniformly integrable in $z$ on a bounded set
$D\subset \C$ if for every $\varepsilon >0$ there exists a $\delta
>0$ such that for every compact set $S\subset D$ of area less than
$\delta$
\[
\int_{S} |h_{\alpha} (z)| d^2z <\varepsilon  \hspace{6ex} \forall
\alpha \in {\cal A}.
\]
It is a corollary of the local integrability of $\log |z|$ that
for every compact set $K\subset \C$ the family of functions $\{
\log |\zeta -z| \}_{\zeta\in K}$ is uniformly integrable in $z$ on
bounded subsets of $\C$. From this one immediately obtains

\bigskip

\begin{Prop} \label{prop3:1} Let $\chi_{R}(|\zeta|)$ be the characteristic function
of the disk $|\zeta|\le R$. For every compact set $K\subset \C$
and for every $R>0$ the family of functions
\[
\Big\{ \frac{1}{n}\sum\limits_{j=1}^{n}\chi_R (|\zeta_j|) \log
|\zeta_j - z |\Big\}_{\!\!n\ge 1, \;  \zeta_1, \ldots , \zeta_n
\in K}
\]
is uniformly integrable in
$z$ on bounded subsets of $\C$.
\end{Prop}

\bigskip

It is now apparent that for every continuous $f$ with compact
support and for every $R>0$
\begin{equation}\label{eps1}
\forall \varepsilon >0 \hspace{2ex} \exists \delta >0:
\hspace{3ex} \limsup_{n\to\infty} \int_{L_{\delta}} \left| f(z)
\frac{1}{n}\sum_{|z_j|\le R}\log |z_j-z|\right| d^2z <
\varepsilon.
\end{equation}

To obtain an appropriate upper bound on the second term in
(\ref{b2}), note the following. If $\zeta$ is such that $\dist
(\zeta, \s J_n) \ge 1$ then $\log |z -\zeta|\ge 0$ for every $z\in
\s J_n$ and
\begin{eqnarray} \label{u1}
0 \le \sum_{|z_j|> R}\log |z_j-\zeta| &\le&  \log |\det (J_n
-\zeta I_n)| \\ \label{u2}  & \le&  \sum_{j=1}^n
 \log \big(e^{\xi_{j-1}} + e^{\eta_j} +|q_j|
+|\zeta|\big).
\end{eqnarray}
The latter inequality is due to the fact that for every matrix
$A=||A_{jk}||_{j,k=1}^n$, $|\det A|\le \prod_{j=1}^n\sum_{k=1}^n
|A_{jk}| $.

By Theorem \ref{th1},  all non-real eigenvalues of $J_n$ are in
the vicinity of $\cal L$ for all sufficiently large $n$. ${\cal L}
$ is a smooth curve and any
vertical line in the upper half of the complex plane intersects
${\cal L}$ only once there. Therefore the probability is one
that, moving up along the imaginary axis say, we can find a
$\zeta_0=iy$ such that $\dist (\zeta_0, \s J_n) \ge 1$ for all
$n\ge n_0$.

Note that
\[
\left|\sum_{|z_j|>R} \log |z_j-z | - \sum_{|z_j|>R} \log
|z_j-\zeta_0 |\right| \le \sum_{|z_j|>R} \left|\log \left|1 +
\frac{\zeta_0-z}{z_j-\zeta_0} \right|\right|
\]
Choose now $r_0$ so that $r_0 >1$ and the disk $|z|\le r_0$
contains both $\zeta_0$ and the support $K$ of $f$. Set $R=4r_0$.
Then for all $|z|\le r_0$ and for all $z_j$ in the exterior of
$|z|\le R=4r_0$ we have $|\zeta_0-z| \le 2 r_0$, $|z_j - \zeta_0|
\ge 3 r_0$ and $|z_j -z| \ge 3 r_0$ and hence
\begin{eqnarray*}
 \left| \frac{1}{n} \sum_{|z_j|>R} \log |z_j-z |\right| &\le& \log 3
+ \left| \frac{1}{n} \sum_{|z_j|>R} \log |z_j-\zeta_0|  \right|
\hspace{18ex} \forall\ |z| \le r_0 \\
   &\le & \log 3 + \frac{1}{n}\sum_{j=1}^n
 \log \big(e^{\xi_{j-1}} + e^{\eta_j} +|q_j|
+r_0\big) \hspace{5ex}\hbox{[by (\ref{u1} -- (\ref{u2}))]}.
\end{eqnarray*}
By the ergodic theorem, under assumptions (\ref{ba1}) --
(\ref{ba2}), on a set $\Omega_1$ of full probability measure,
\[
\lim_{n\to\infty} \frac{1}{n}\sum_{j=1}^n
 \log \big(e^{\xi_{j-1}} + e^{\eta_j} +|q_j|
+r_0\big) = \E \log \big(e^{\xi_{0}} + e^{\eta_1} +|q_1| +r_0\big)
< +\infty
\]
Therefore, on $\Omega_1$,
\[
\limsup_{n\to\infty} \left\{ \frac{1}{n} \sup_{|z|\le
r_0}\left|\sum_{|z_j|>R} \log |z_j-z |\right|\right\} \le Const
\]
and, because the disk $|z|\le r_0$  covers   the support of $f$,
\begin{equation}\label{eps2}
\forall \varepsilon >0 \hspace{2ex} \exists \delta >0:
\hspace{3ex} \limsup_{n\to\infty} \int_{L_{\delta}} \left| f(z)
\frac{1}{n}\sum_{|z_j|> R}\log |z_j-z|\right| d^2z < \varepsilon.
\end{equation}

(\ref{eps1}) and (\ref{eps2}) imply (\ref{eps}). Lemma \ref{sec3:th2}
 is proved. \hfill $\blacksquare$

\bigskip


\bigskip

\begin{Corol}\label{sec3:th4}
Assume (\ref{ba1}) -- (\ref{ba2}). Then with probability one,
\[
\lim_{n\to\infty} d\nu_{\!\scriptscriptstyle{J_n}}=d\nu
\]
in the sense of weak convergence of measures.
\end{Corol}

\bigskip

\noindent \emph{Proof}. Since the operation $\Delta$ is continuous
on distributions, Lemma \ref{sec3:th2} implies that, on a set of
full probability measure,
\[
d\nu_{\!\scriptscriptstyle{J_n}}\! (z)=\frac{1}{2\pi}\Delta
p(z;J_n) \to \frac{1}{2\pi}\Delta p(z)=d\nu (z), \hbox { as }
n\to\infty,
\]
as distributions. To complete the proof, recall that a sequence of
measures converging as distributions must be converging weakly
\cite{Hoermander1}. \hfill $\blacksquare$

\bigskip

It is apparent that $\Delta p(z)=0$ everywhere off a line
consisting of two parts. One is the equipotential line $\cal L$
(\ref{0018}) that separates the domains $D_1$ and $D_2$; the
other is $\Sigma$ (\ref{Sigma}) which is made up of all points of
$\Supp dN$ that do not belong to the interior of the closed
contours of ${\cal L}$.

$p(z)$ is continuous in the upper and lower parts of the complex
plane. It also has continuous derivatives everywhere but on $\cal
L$ and $\Sigma$. Its normal derivative has a jump when $z$ moves
from $D_2$ to $D_1$ in the direction perpendicular to $\cal L$. It
follows from this that the restriction $d\nu^C$ of $d\nu$ to
$\C\backslash\R$ is supported on $\cal L$ and has there density
$\rho (z)$ with respect to the arc-length measure $dl$ on $\cal
L$. The density equals the jump in the normal derivative of $p(z)$
multiplied by $\frac{1}{2\pi}$. A straightforward calculation
gives
\[
\rho
(z)=\frac{1}{2\pi}\left|\int_{-\infty}^{+\infty}\frac{dN(\lambda)}{\lambda
-z} \right|.
\]
On the other hand, the restriction $d\nu^R$ of $d\nu $ to the real
axis is supported on $\Sigma$ and coincides there with $dN$.
Therefore, for every bounded continuous function $f(z)$
\begin{eqnarray}\label{aa1}
\int_{\C} f(z) d\nu (z) &=& \int_{\C} f(z) d\nu^R(z) + \int_{\C}
f(z) d\nu^C(z)
 \\ &=& \label{aa2} \int_{\Sigma} f(\lambda) dN(\lambda) +
\int_{\cal L} f(z(l)) \rho (z(l)) dl.
\end{eqnarray}
Taking into account that the weak convergence of measures $d\nu_n$
is equivalent to the convergence of $\int f(z) d\nu_n(z)$ on
bounded continuous functions we obtain from (\ref{aa1}) --
(\ref{aa2}) and Corollary {\ref{sec3:th4} the statement of Theorem
\ref{th2}.

\appendix
\section{Appendix}
\label{appendix}

In our analysis of the non-self-adjoint eigenvalue problem
(\ref{4})-(\ref{5}) we have used a number of results about the
finite difference equation
\begin{equation}\label{2.1}
-c_{k-1}\varphi_{k-1} - c_{k} \varphi_{k+1}
+q_{k}\varphi_{k}=z\varphi_{k}
\end{equation}
with random coefficients $c_k$ and $q_k$. For the sake of
completeness, we reproduce here the corresponding formal
statements. With the exception of Theorem \ref{Th2.5b} these
results are well known in the theory of random selfadjoint
operators. Their proofs together with references to the original
publications can be found in books \cite{CL,PF}. Some of the
results are proved there under slightly less general assumptions
than those used in this paper. However only minor adjustments are
needed to extend the published proofs to the generality of our
assumptions.

Given a sequence of 2-component vectors $\{(c_k,q_k)
\}_{j=1}^{\infty} $,  consider a sequence of tridiagonal symmetric
matrices $H_n$ of dimension $n$, $n=1,2,\ldots , $ defined on
$\{(c_k,q_k)\}_{j=1}^{\infty} $ in the following way
\[
H_n = \left(
\begin{array}{c c c c}
q_{1}  & -c_{1}   &            &
\\
-c_{1} &  \hspace{-2ex}\ddots  & \hspace{-2ex}\ddots     &
\\
       &\hspace{-2ex} \ddots    &  \hspace{-2ex} \ddots   &
\hspace{-2ex} -c_{n-1}
\\
       &          &  \hspace{-2ex}-c_{n-1}  & \hspace{-2ex} q_n
\\
\end{array}
\right)
\]

Assume that
\begin{eqnarray}\label{a1}
& &
\begin{array}{l}
\omega\equiv \{ (c_j, q_j) \}_{j=1}^{\infty} \hbox{ is a
stationary ergodic (with respect to the translation $k \to k+1$) }
\\ \hbox{sequence of random vectors defined on a common probability space}\
(\Omega, {\cal F}, P);
\end{array}\hspace{3ex}
\\[1ex]
\label{a2} & &
\begin{array}{l} c_k >0 \hspace{1ex} \hbox{for all $k$, and }\hspace{1ex}
\E \ln (1+|q_1|) < +\infty,\ \E |\ln c_1| < +\infty .
\end{array}
\end{eqnarray}
Here as before the symbol $\E$ denotes averaging ever the
probability space.

The matrix $H_n$ is symmetric and has real eigenvalues. Their
empirical cumulative distribution function is defined as
\[
N_n(\lambda, \omega )=\frac{1}{n}\# \{\hbox{eigenvalues of $H_n$
in $(-\infty, \lambda)$} \}.
\]

\bigskip

\begin{Th}
\label{Th2.1} Assume (\ref{a1}) -- (\ref{a2}). Then there exists a
continuous non-random function $N(\lambda)$ of real variable
$\lambda$ such that for almost all sequences $\omega $
\[
\lim_{n\to\infty} N_n(\lambda, \omega) = N(\lambda).
\]
In other words, on a set of full probability measure, the
eigenvalue counting measures $dN_n(\lambda, \omega )$ converge
weakly, as $n\to\infty$, to the limiting measure $dN(\lambda)$.
\end{Th}

The limit distribution function, $N(\lambda )$, is called the
integrated density of states (IDS) of Eq.\ (\ref{2.1}).

Let
\begin{equation}\label{2.5}
p(z, H_n) = \int_{-\infty}^{+\infty}\log |\lambda -z|\
dN_n(\lambda, \omega )
\end{equation}
and
\begin{equation}\label{2.8}
\Phi(z)
=\int_{-\infty}^{+\infty}\log |\lambda -z|\
dN(\lambda ).
\end{equation}
Under assumptions (\ref{a1}) -- (\ref{a2}), the integral in
(\ref{2.8}) converges for every non-real $z$, hence $\Phi (z)$ is
well defined off the real axis. On the real axis, the equality in
(\ref{2.8}) is understood in the following sense\footnote{Under
assumptions (\ref{a1}) -- (\ref{a2}), the prelimit integral
converges for every $\varepsilon >0$ }
\[
\Phi (x)= \lim_{\varepsilon \downarrow 0}
\int_{-\infty}^{+\infty}\max [\log |\lambda - x|, -\varepsilon^{-1} ]\
dN(\lambda ), \hspace{5ex} x\in \R
\]
with the convention that $\Phi(x)=-\infty$ if the above
limit is $-\infty$\footnote{In fact, $\Phi (z)$ is finite
everywhere in the complex plane, see (\ref{final}).}. With this
convection, $\Phi (z) $ is subharmonic in the complex plane. In
particular, $\Phi (z)$ is upper semi-continuous.

\bigskip

\begin{Th}
\label{Th2.3} Assume (\ref{a1}) -- (\ref{a2}). Then the following
is true for  almost all $\omega $: For all non-real $z$
\begin{equation}\label{2.13}
\lim_{n\to \infty} p(z; H_n) = \Phi (z).
\end{equation}
The convergence in
(\ref{2.13}) is uniform in $z$ on every compact subset of
$\C \backslash \R $.
\end{Th}

\smallskip

Let
\begin{equation}\label{2.17}
\gamma_n(z, \omega)= \frac{1}{n} \log || S_n (z)||,
\end{equation}
where $S_n(z)$ is as in (\ref{tranmat}).

The Lyapunov exponent is defined as follows:
\[
\bar \gamma (z)= \lim_{n\to\infty} \E \gamma_n(z, \omega).
\]
The limit above exists for every $z\in \C$. Any matrix norm can be
used in (\ref{2.17}), as they all are equivalent. Since
$||S_n||^2\ge |\det S_n|$ and  $\det S_n (z)=c_0/c_n$, we have
that for every $z\in \C$, and in particular for every real $z$,
\[
\bar \gamma (z) \ge 0.
\]

\bigskip

\begin{Th}
\label{Th2.5}  Assume (\ref{a1}) -- (\ref{a2}). Then for almost
all $\omega$ the following is true: For every compact set $K
\subset \C\backslash \R$
\[
\lim_{n \to \infty} \gamma_n (z, \omega) = \bar \gamma (z)
\hspace{4ex} \hbox{uniformly in $z\in K$}.
\]
\end{Th}

\bigskip

\noindent {\bf Remark}. In contrast to non-real $z$, on the real
axis
\begin{equation}\label{limg}
\lim_{n \to \infty} \gamma_n (x,\omega) = \gamma (x,\omega),
\end{equation}
The limit above exists for almost all pairs $(x,\omega)$ and $\E
\gamma (x,\omega) = \bar \gamma (z)$. However, when $x$ is fixed,
the set $\Omega_x$ of those $\omega$ for which the limit in
(\ref{2.17}) exists depends on $x$, and $P(\cap_{x\in \Sigma }
\Omega_x )=0$ (see \cite{G,AS}).

\smallskip

For every $\omega$, $\gamma_n (z, \omega)$ is a subharmonic
function in the complex plane. $\bar \gamma (z)$ is also
subharmonic in $\C$. This property of the Lyapunov exponent is
very useful, see \cite{CS}. We use it here to deduce the following
corollary from Theorem \ref{Th2.5}.

\bigskip

\begin{Th}
\label{Th2.5b}  Assume (\ref{a1}) -- (\ref{a2}). Then for almost
all $\omega$ the following is true: For every compact set
$K\subset \C$, and in particular for every compact set $K\subset
\R$,
\[
\limsup_{n\to \infty}\left\{\sup_{z\in K} \gamma_n(z,
\omega)\right\}  \le \sup_{z\in K} \bar \gamma (z).
\]
If, in addition, $\bar \gamma (z) $ is continuous in $K$ then
\[
\limsup_{n\to \infty}\left\{\sup_{z\in K} [\gamma_n(z, \omega)-
\bar \gamma (z)] \right\}  \le 0.
\]

\end{Th}

\bigskip

\noindent {\bf Remark}. This theorem plays a crucial role in our
proof of the fact that the eigenvalues of $J_n$ are wiped out, as
$n\to\infty $, from the interior of each contour of the curve
$\cal L$, see part (b) of Theorem \ref{th1}. Actually, our proof
of part (a) of Theorem \ref{th1} also applies to any compact
subset of $D_2\backslash\R$. However, this proof is based on
Theorem \ref{Th2.5} and cannot be applied to the whole interior of
$\cal L$ as it contains intervals of real axis.

\bigskip

Theorem \ref{Th2.5b} follows immediately from Theorem \ref{Th2.5}
and the following result from the theory of subharmonic functions:

\medskip
\begin{Th}\label{ThH}[see \cite{H}, p.\ 150]
Let $u_j\equiv \hspace{-2ex}/ \hspace{0.5ex}-\infty$ be a sequence
of subharmonic functions in $\C$ converging in the sense of
distribution theory to the subharmonic function $u$. If $K$
is a compact subset of $\C$ and $f$ is continuous on $K$, then
\[
\limsup_{j\to \infty}\left\{\sup_{K} (u_j- f) \right\}
\le\sup_{K}(u-f).
\]
\end{Th}

\bigskip

\begin{Th}[Thouless formula]
\label{Th2.6}  For all $z\in \C$, $ \bar \gamma (z)= \Phi (z) - \E
\log c_1 $.
\end{Th}

\bigskip

It is a corollary of the Thouless formula and the positivity of
the Lyapunov exponent $\bar \gamma (z)$ that
\begin{equation}\label{final}
\Phi (z) \ge \E\log c_0  \hspace{10ex} \forall z \in\C.
\end{equation}

\end{document}